\documentclass[12pt]{article}
\usepackage{amsfonts}
\usepackage{palatino}
\usepackage{textcomp}
\usepackage{color}
\usepackage{amsmath}
\usepackage{graphicx}
\usepackage{amssymb}
\usepackage{dsfont}
\usepackage{float}
\usepackage{epstopdf}

\usepackage{authblk}
\begin{document}



\def\a{\alpha}
\def\b{\beta}
\def\d{\delta}
\def\e{\epsilon}
\def\g{\gamma}
\def\h{\mathfrak{h}}
\def\k{\kappa}
\def\l{\lambda}
\def\o{\omega}
\def\p{\wp}
\def\r{\rho}
\def\t{\tau}
\def\s{\sigma}
\def\z{\zeta}
\def\x{\xi}
\def\V={{{\bf\rm{V}}}}
 \def\A{{\cal{A}}}
 \def\B{{\cal{B}}}
 \def\C{{\cal{C}}}
 \def\D{{\cal{D}}}
\def\K{{\cal{K}}}
\def\O{\Omega}
\def\R{\bar{R}}
\def\T{{\cal{T}}}
\def\L{\Lambda}
\def\f{E_{\tau,\eta}(sl_2)}
\def\E{E_{\tau,\eta}(sl_n)}
\def\Zb{\mathbb{Z}}
\def\Cb{\mathbb{C}}

\newcommand{\vev}[1]{{\left< {#1} \right>}}
\newcommand{\bra}[1]{{\left< {#1} \right|}}
\newcommand{\ket}[1]{{\left| {#1} \right>}}

\def\R{\overline{R}}

\def\beq{\begin{equation}}
\def\eeq{\end{equation}}
\def\bea{\begin{eqnarray}}
\def\eea{\end{eqnarray}}
\def\ba{\begin{array}}
\def\ea{\end{array}}
\def\no{\nonumber}
\def\le{\langle}
\def\re{\rangle}
\def\lt{\left}
\def\rt{\right}

\newtheorem{Theorem}{Theorem}
\newtheorem{Definition}{Definition}
\newtheorem{Proposition}{Proposition}
\newtheorem{Lemma}{Lemma}
\newtheorem{Corollary}{Corollary}
\newcommand{\proof}[1]{{\bf Proof. }
        #1\begin{flushright}$\Box$\end{flushright}}

\title{Shor-Movassagh chain leads to unusual integrable model}
\author[1]{\large \bf Bin Tong}
\author[2]{\large \bf Olof Salberger}
\author[1,2]{\large \bf Kun Hao\footnote{haoke72@163.com}}
\author[2]{\large \bf Vladimir Korepin}
\affil[1]{\normalsize Institute of Modern Physics, Northwest University, Xi'an 710127, China}
\affil[2]{\normalsize C.N. Yang Institute for Theoretical Physics, Stony Brook University, NY 11794, USA}

\maketitle
The ground state of Shor-Movassagh chain can be analytically  described by the Motzkin paths. There is no analytical description of the  excited states, the model is not solvable.
We prove the integrability of the model without  interacting part in this paper [free Shor-Movassagh].
The Lax pair for the free Shor-Movassagh open chain is explicitly constructed.
We further obtain the boundary $K$-matrices compatible with the integrability of the model on the open interval.
Our construction provides a direct demonstration for the quantum  integrability of the model, described by Yang-Baxter algebra.
Due to the lack of crossing unitarity, the integrable open chain can not be constructed by the reflection equation (boundary Yang-Baxter equation).

\section{Introduction}
\setcounter{equation}{0}
A  spin-$1$ model called  Shor-Movassagh chain \cite{shor}  was shown to have unique ground state \footnote{Its half-integer version is called the Fredkin spin chain \cite{shor2,fredkin}.}.  The ground state can be expressed as a sum with respect to Motzkin paths.
Among other interesting aspects, its entanglement entropy strongly exceeds the one exhibited by other previously known local models.
Inspired by the discovery of the Affleck-Kennedy-Lieb-Tasaki (AKLT) model, studying such Hamiltonians with highly entangled spins is currently one of the most challenging and intriguing fields in quantum physics.
Since this model is not exactly solvable, it is still hard to describe thermodynamics.
The model can be seen as a generalized Temperley-Lieb system.
This is understood from the local Hamiltonian, since it is a representation of the generator of Temperley-Lieb algebra.

\begin{figure}[H]
\begin{center}
		\includegraphics[height=1.7in]{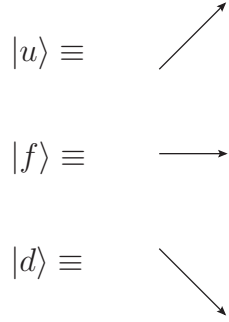}
	\caption{\small The local Hilbert space and its mapping to the steps in the ``x-y" plane. }
\label{ufd}
\end{center}
\end{figure}

The full Hamiltonian of the uncolored\footnote{The color was introduced  by Shor \& Movassagh \cite{shor}.} Shor-Movassagh chain was described by Bravyi et al. \cite{Bravyi2012}. It is a spin chain with a 3-dimensional local Hilbert space given by a basis $\ket{u}$, $\ket{f}$, and $\ket{d}$. Here $u$, $f$, and $d$ are used to abbreviate ``up'', ``flat'', and ``down'' of the Motzkin walks respectively.
This identification comes from that the states in this system can be mapped to paths (Motzkin walks) in the ``x-y" plane as done in \cite{shor}.
So the state $\ket{u}$ maps to the $(1,1)$ direction, $\ket{f}$ maps to the $(1,0)$ direction, and $\ket{d}$ maps to the $(1,-1)$ direction in the ``x-y" plane, as shown in figure \ref{ufd}.

The hamiltonian is given by (the coupling constant $g$ was equal to 1 in the original paper):
\[
 H_{\text{SM}} = H_{boundary}+\sum_{j=1}^{N-1} H_{free,j} + g H_{int,j}
\]
In the original paper the Hamiltonian densities $H_{free}$ and $H_{int}$ are given in terms of two-site projectors, as
\[
H_{free,j} = (\ket{u_jf_{j+1}} - \ket{f_ju_{j+1}}) (\bra{u_jf_{j+1}} - \bra{f_ju_{j+1}}) + (\ket{d_jf_{j+1}} - \ket{f_jd_{j+1}})(\bra{d_jf_{j+1}} - \bra{f_jd_{j+1}})\]
and
\[H_{int,j} =(\ket{u_jd_{j+1}} - \ket{f_jf_{j+1}}) (\bra{u_jd_{j+1}} - \bra{f_jf_{j+1}}).
\]
The boundary operators for the open chain are given by
\[
H_{boundary}=H_1+H_N=\ket{d_1}\bra{d_1}+\ket{u_N}\bra{u_N}.
\]
In this paper we will analyze the case where the coupling contacnt $g=0$.
We  observe that the resulting system is integrable. By turning on $g$, the system loses its symmetries and
starts to ``interact". The corresponding term can be viewed as a ``kinetic part" of the Shor-Movassagh model.
Thus we will call the case with $g=0$ as the non-interacting Shor-Movassagh
chain or the free Shor-Movassagh spin chain. We will denote the corresponding Hamiltonian $H_{FSM,open}$ in the following.

To readers familiar with other similar models but not with the Bethe ansatz, the central building block is the $R$-matrix (\ref{R-matrix-Motzkin}). It essentially describes the scattering between two quasiparticles in the spin chain. This $R$-matrix satisfies the Yang-Baxter equation (\ref{YBE}), which states that many-body scattering can be reduced to two-body scattering. This fact ensures the integrability of the periodic free Shor-Movassagh chain.

The bulk of the Hamiltonian $H_{FSM,open}$ (\ref{Hamiltonian}) can be derived from the periodic transfer matrix (constructed by $R$-matrix) through a logarithmic trace identity. However, the open boundary case Hamiltonian $H_{FSM,open}$ can not be constructed by the double row transfer matrix based on Sklyanin's reflection equations \cite{Sklyanin88} or the boundary $S$ matrices (for particles scattering) \cite{Fendley-Saleur94}. This is because the partial transpose of this $R$-matrix is degenerate (\ref{partial-trans}), the $R$-matrix does not have crossing unitarity \cite{Cao13,Wang15}. One can not construct the open chain in the ordinary way.

On the other hand, the traditional basis for applying the quantum inverse scattering method (QISM) to a completely integrable system is to represent the equations of motion of the system into the Lax form \cite{Lax68,Lax80}. For quantum systems with periodic boundary conditions, the existence of the $R$-matrix allows one to express the original equations of motion in the quantum Lax form \cite{Korepin82,Korepin93,Takhtajan87}. Meanwhile the explicit forms of Lax pairs for some physically important models have been given by many authors \cite{Wadati1982,Wadati1987,Zhou1989}.
A natural problem is that there must exist one kind of revised form of the ordinary Lax formulation for models with open boundaries. It can be used to describe completely quantum integrable lattice spin open chains \cite{Zhou1996}.

The aim of this paper is to present an explicit construction of the Lax pair for the free Shor-Movassagh open chain.
The construction provides a direct demonstration for the quantum
integrability of the model. As a further result, we obtain its boundary $K$-matrices
compatible with the integrability of the model. We relate this random walk model to a integrable spin chain.

In this paper we will employ Lax formulation to approach the problem of the free Shor-Movassagh chain for open boundary case.
The paper is organized as follows. In section 2, the Lax formulation for open quantum spin chain is introduced.
In section 3, the Hamiltonian of free Shor-Movassagh chain with open boundary conditions is given in more detailed form.
The Lax pair for the open free Shor-Movassagh chain is explicitly constructed in section 4. The corresponding boundary $K$-matrices are obtained in section 5. Section 6 is devoted to the conclusion.

\section{Lax pair}
\setcounter{equation}{0}
Due to the $R$-matrix of free Shor-Movassagh chain does not have the crossing unitary. Sklyanin's reflection equations \cite{Sklyanin88} do not apply to this case. We try to construct reflection equations by using quantum Lax formulation.

We first recall the original work of Lax formulation \cite{Korepin82,Korepin93,Takhtajan87} for quantum integrable models with general open boundary conditions in one dimension \cite{Zhou96,Guan97,Guan2000}.  We consider an operator version of the auxiliary linear problem
\begin{eqnarray}
\Phi_{j+1} & = & L_j(\lambda)\Phi_j,\quad j=1,2,\ldots ,N,\nonumber\\
\frac{d}{dt}\Phi_j & = & M_j(\lambda)\Phi_j,\quad j=2,3,\ldots ,N,\label{equ-motion}
\end{eqnarray}
with boundary equations
\begin{eqnarray}
\frac{d}{dt}\Phi_{N+1} & = & M_+(\lambda)\Phi _{N+1},\nonumber\\
\frac{d}{dt}\Phi_1 & = & M_-(\lambda)\Phi_1.\label{equ-motion-boundary}
\end{eqnarray}
Here $L_j(\lambda)$, $M_j(\lambda)$, and $M_{\pm}(\lambda)$ are matrices\footnote{We remark that the matrix $L_j$ acts on auxiliary space $0$ and quantum space $j$. $M_j$ acts on spaces $0$, $j-1$, and $j$. $M_{-}$ ($M_{+}$) acts on spaces $0$ and $1$ ($N$), respectively. We will give further details in later sections.}  depending on
the spectral parameter $\lambda$ and the dynamical variables. The spectral parameter $\lambda$ does not depend on the time $t$, and dynamical variables. Evidently, the consistency conditions
for Eqs.~(\ref{equ-motion}) and (\ref{equ-motion-boundary}) yield the following Lax equations
\begin{equation}
\frac{d}{dt}L_j(\lambda)=M_{j+1}(\lambda)L_j(\lambda)-L_j(\lambda)M_j(\lambda),
\quad j=2,3,\ldots ,N-1,\label{Lax-bulk}
\end{equation}
and boundary terms
\begin{eqnarray}
\frac{d}{dt}L_N(\lambda)&=&M_+(\lambda)L_N(\lambda)-L_N(\lambda)M_N(\lambda),\nonumber\\
\frac{d}{dt}L_1(\lambda)&=&M_2(\lambda)L_1(\lambda)-L_1(\lambda)M_-(\lambda).\label{Lax-boundary}
\end{eqnarray}
If the equations of motion for the system can be expressed in the form of equations (\ref{Lax-bulk}) and (\ref{Lax-boundary}), with the boundary $K$-matrices exist as the solutions of equations (\ref{K-reflection}) and (\ref{K+reflection}) below, then we insist that the spin chain with open boundary conditions is completely integrable.
Let $T(\lambda)=L_N(\lambda)\cdots L_2(\lambda) L_1(\lambda)$ be the usual monodromy matrix
\cite{Korepin93} of the system. $K_{-}(\lambda)$ and $K_{+}(\lambda)$ are the matrices for
the left and the right boundary, respectively. The double row transfer matrix
$\tau(\lambda)$ of open chain is defined as the trace on the auxiliary space $V_{0}$ as
\begin{equation}
\tau(\lambda)={\rm tr}_0\left[K_+(\lambda)T(\lambda)K_-(\lambda)T^{-1}(-\lambda)\right].\label{Double-trans}
\end{equation}
From the Lax equations (\ref{Lax-bulk}) and (\ref{Lax-boundary}), it follows that
the transfer matrix $\tau(\lambda)$ does not depend on time $t$. The
boundary matrices satisfy the conditions
\begin{eqnarray}
K_-(\lambda)M_-(-\lambda) & = & M_-(\lambda)K_-(\lambda),\label{K-reflection}\\
{\rm tr}_0\left[K_+(\lambda)M_+(\lambda)\mathbb{U}_N(\lambda)\right] & = &
{\rm tr}_0\left[K_+(\lambda)\mathbb{U}_N(\lambda)M_+(-\lambda)\right],\label{K+reflection}
\end{eqnarray}
where $\mathbb{U}_N(\lambda)=T(\lambda)K_{-}(\lambda)T^{-1}(-\lambda)$.  This implies that the double row transfer matrices with different spectral parameters commute with each other,
\begin{equation}
[\tau(\lambda),\tau(\mu)]=0.
\end{equation}
Thus the
system possesses an infinite number of independent conserved
quantities, and it is completely integrable.

\section{The free Shor-Movassagh chain with open boundary conditions}
\setcounter{equation}{0}

We denote the basis states by $\{\left|{u}\right>, \ket{f}, \ket{d}\}$, where $u$, $f$ and $d$ are used to abbreviate ``up", ``flat", and ``down" of Motzkin walks, respectively,
\begin{equation}
\ket{u}=\left(\begin{array}{c}
    1 \\
    0 \\
    0 \\
  \end{array}\right),\quad
\ket{f}=\left(\begin{array}{c}
    0 \\
    1 \\
    0 \\
  \end{array}\right),\quad
\ket{d}=\left(\begin{array}{c}
    0 \\
    0 \\
    1 \\
  \end{array}\right).
\end{equation}

Then the Hamiltonian for the open free Shor-Movassagh chain can be expressed as
\begin{equation}
H_{FSM, open}=H_1+\sum^{N-1}_{j=1}[\hat{U}_{j,j+1}+\hat{D}_{j,j+1}]+H_N.
\label{Hamiltonian}\end{equation}
The operators $\hat{U}_{j,j+1}$ and $\hat{D}_{j,j+1}$ are projectors to the states
$$ \ket{u_j,f_{j+1}} - \ket{f_j,u_{j+1}}, $$
and
$$ \ket{d_j,f_{j+1}} - \ket{f_j,d_{j+1}} $$
respectively.

Then the operator $\hat{U}_{j,j+1}$ can be expressed in terms of the states and standard basis,
\begin{eqnarray}
&&\ket{u_j,f_{j+1}}\bra{u_j,f_{j+1}} - \ket{u_j,f_{j+1}}\bra{f_j,u_{j+1}} - \ket{f_j,u_{j+1}}\bra{u_j,f_{j+1}} + \ket{f_j,u_{j+1}}\bra{f_j,u_{j+1}},\no\\
&=&E_j^{11}E_{j+1}^{22}-E_j^{12}E_{j+1}^{21}-E_j^{21}E_{j+1}^{12}+E_j^{22}E_{j+1}^{11},
\end{eqnarray}
and for the operator $\hat{D}_{j,j+1}$,
\begin{eqnarray}
&&\ket{d_j,f_{j+1}}\bra{d_j,f_{j+1}} - \ket{d_j,f_{j+1}}\bra{f_j,d_{j+1}} - \ket{f_j,d_{j+1}}\bra{d_j,f_{j+1}} + \ket{f_j,d_{j+1}}\bra{f_j,d_{j+1}}\no\\
&=&E_j^{33}E_{j+1}^{22}-E_j^{32}E_{j+1}^{23}-E_j^{23}E_{j+1}^{32}+E_j^{22}E_{j+1}^{33}.
\end{eqnarray}

Let us consider the boundary terms
\begin{eqnarray}
H_1&=&a_1|u_1\rangle\langle u_1|+b_1|f_1\rangle\langle f_1|+c_1|d_1\rangle\langle d_1|
=a_1E_1^{11}+b_1E_1^{22}+c_1E_1^{33},\\
H_N&=&a_N|u_N\rangle\langle u_N|+b_N|f_N\rangle\langle f_N|+c_N|d_N\rangle\langle d_N|
=a_{N}E_N^{11}+b_{N}E_N^{22}+c_{N}E_N^{33}.
\end{eqnarray}

We remark that this Hamiltonian with open boundaries can not be derived from Sklyanin's reflection equations \cite{Sklyanin88}. In the following sections, we will construct the Hamiltonian (\ref{Hamiltonian}) based on Lax formulation, and its integrability will be discussed.

\section{Constructing the Lax pair operators for bulk and boundaries}
\setcounter{equation}{0}
Unless specified otherwise, here and below we adopt the standard notation used in Algebraic Bethe Ansatz method: for any matrix $A\in {\rm End}({\rm\bf V})$, $A_j$ is an embedding operator in the tensor space ${\rm\bf V}\otimes
{\rm\bf V}\otimes\cdots$, which acts as $A$ on the $j$-th space and as an identity on the other factor spaces.

The $R$-matrix $R_{12}(\lambda)=P_{12}[(\lambda+\eta)I-\lambda\hat{e}_{12}]$ for the free Shor-Movassagh spin chain can be written as
\begin{equation}
\label{R-matrix-Motzkin}
R_{12}(\lambda)=
\left(
\begin{array}{ccc|ccc|ccc}
\lambda+\eta &    &    &    &    &    &    &    &   \\
      &\lambda&    &\eta&    &    &    &    &   \\
      &    &    &    &    &    &\lambda+\eta&    &   \\
\hline
      &\eta&    &\lambda&    &    &    &    &   \\
      &    &    &    &\lambda+\eta&    &    &    &   \\
      &    &    &    &    &\lambda&    &\eta&   \\
\hline
      &    &\lambda+\eta&    &    &    &    &    &    \\
      &    &    &    &    &\eta&    &\lambda&    \\
      &    &    &    &    &    &    &    &\lambda+\eta\\
\end{array}
\right).
\end{equation}
Here $\lambda$ is the spectral parameter. $\eta$ is the crossing parameter. $P_{12}$ is the permutation operator, $\hat{e}_{12}$ is the corresponding generator \cite{Temperley-Lieb} of the Temperley-Lieb algebra, and ${1\over2}\hat{e}_{12}$ is a projection operator.
\begin{equation}
\hat{e}_{12}=\hat{U}_{12}+\hat{D}_{12}.
\end{equation}
Here we give some discussions about this generator $\hat{e}_{12}$:
\begin{eqnarray}
\hat{e}_{j,j+1}\,\hat{e}_{j+1,j+2}\,\hat{e}_{j,j+1} & = & \hat{e}_{j,j+1}, \label{e1}\\
\hat{e}_{j,j+1}\,\hat{e}_{j-1,j}\,\hat{e}_{j,j+1} & = & \hat{e}_{j,j+1}, \label{e2}\\
\hat{e}_{j,j+1}^2 & = & 2\hat{e}_{j,j+1}, \\
\hat{e}_{j,j+1}\,\hat{e}_{k,k+1} & = & \hat{e}_{k,k+1}\hat{e}_{j,j+1},~~|j-k|>1\label{e4}.
\end{eqnarray}
These are the relations for the generators of the Temperley-Lieb algebra. And it is known that the XXX spin chain can be realized using these generators \cite{Saleur89,Martin-Saleur94,Nichols06}.
Comparing these relations to the standard definition of the Temperley-Lieb algebra given in \cite{Nichols06}
\begin{equation}
\hat{e}_{i,i+1}^2 = (q+q^{-1})\hat{e}_{i,i+1}.
\end{equation}
With the other relations being the same as in equations (\ref{e1}), (\ref{e2}), and (\ref{e4}), we find that $q=1$ in our case.\footnote{This confirms that the free Shor-Movassagh spin chain with periodic boundary conditions is integrable from the point view of both the Yang-Baxter algebra and Temperley-Lieb algebra.
We further remark that the discussions for open boundary cases in Refs. \cite{Nichols06,Nichols09,Kulish11} can not be apply to free Shor-Movassagh case.
Since the Temperley-Lieb algebra related $R$-matrix in there satisfies the crossing unitarity. And in their papers the two boundary Temperley-Lieb algebra is related to Sklyanin's reflection equations.}

$R_{ij}(\lambda)$ is an embedding operator of $R$-matrix in the tensor space, which acts as an identity on the factor spaces except for the $i$-th and $j$-th ones.
The $R$-matrix satisfies the quantum Yang-Baxter equation
\begin{equation}
R_{12}(\lambda)R_{13}(\lambda+\nu)R_{23}(\nu)=R_{23}(\nu)R_{13}(\lambda+\nu)R_{12}(\lambda),
\label{YBE}\end{equation}
and possesses the following properties:
\begin{eqnarray}
\mbox{Initial condition:}& R_{12}(0)= \eta P_{12},\label{Initial-R}\\
\mbox{Unitarity:} & R_{12}(\lambda)R_{21}(-\lambda)= (\eta+\lambda)(\eta-\lambda)\,{\rm id},\label{Unitarity}\\
\mbox{Projection:} & R_{12}(-\eta)= \eta P_{12}\hat{e}_{12}= -\eta\hat{e}_{12}.\label{Projection}
\end{eqnarray}
The partial transpose of this $R$-matrix is degenerate
\begin{equation}
\det(R^{t_1}_{12})=0,
\label{partial-trans}
\end{equation}
so it does not satisfy crossing unitarity.

In the case of free Shor-Movassagh chain, the $R$-matrix (\ref{R-matrix-Motzkin}) can be chosen as the $L$-matrix.
Following the method in the refs \cite{Wadati1982,Wadati1987,Zhang1991}, we can calculate the corresponding $M$-matrix\footnote{We point out that the matrix $M_j(\lambda)$ acts non-trivially on spaces $0$, $j-1$, and $j$. Here we only use $j$ as subscript of $M$ is to follow the conventions in the previous papers mentioned above. $M_j(\lambda)$ is the operator for the bulk (of the open boundary case) and also for the periodic case.} $M_j(\lambda)$ in the bulk,
\begin{eqnarray}
M_j(\lambda)=\left(
\begin{array}{ccc}
M_{j-1,j}^{11}&M_{j-1,j}^{12}&M_{j-1,j}^{13}\\
M_{j-1,j}^{21}&M_{j-1,j}^{22}&M_{j-1,j}^{23}\\
M_{j-1,j}^{31}&M_{j-1,j}^{32}&M_{j-1,j}^{33}\\
\end{array}
\right),\qquad j=2,\cdots,N,
\end{eqnarray}
in which
\begin{eqnarray}
&&M_{j-1,j}^{11}=\frac{i\eta^2}{\eta^2-\lambda^2}(\mathds{1}-2E_{j-1}^{22}E_{j}^{22})
-\frac{i\eta}{\eta-\lambda}E_{j-1}^{12}E_{j}^{21}
-\frac{i\eta}{\eta+\lambda}E_{j-1}^{21}E_{j}^{12}\no\\
&&~~~~~~~~~~~~~~~~~~~-i E_{j-1}^{23}E_{j}^{32}-i E_{j-1}^{32}E_{j}^{23},\no\\
&&M_{j-1,j}^{22}=\frac{i\eta^2}{\eta^2-\lambda^2}[\mathds{1}-2(E_{j-1}^{11}
+E_{j-1}^{33})(E_{j}^{11}+E_{j}^{33})]-\frac{i\eta}{\eta+\lambda}(E_{j-1}^{12}E_{j}^{21}+E_{j-1}^{32}E_{j}^{23})\no\\
&&~~~~~~~~~~~~~~~~~~~-\frac{i\eta}{\eta-\lambda}(E_{j-1}^{21}E_{j}^{12}+E_{j-1}^{23}E_{j}^{32}),\no\\
&&M_{j-1,j}^{33}=\frac{i\eta^2}{\eta^2-\lambda^2}(\mathds{1}-2E_{j-1}^{22}E_{j}^{22})
-\frac{i\eta}{\eta-\lambda}E_{j-1}^{32}E_{j}^{23}
-\frac{i\eta}{\eta+\lambda}E_{j-1}^{23}E_{j}^{32}\no\\
&&~~~~~~~~~~~~~~~~~~~-i E_{j-1}^{21}E_{j}^{12}-i E_{j-1}^{12}E_{j}^{21},\no\\
&&M_{j-1,j}^{12}=\frac{i\eta}{\eta-\lambda}(E_{j-1}^{11}E_{j}^{21}+E_{j-1}^{21}E_{j}^{22}
+E_{j-1}^{33}E_{j}^{21})
+\frac{i\eta}{\eta+\lambda}[E_{j-1}^{21}(E_{j}^{11}+E_{j}^{33})+E_{j-1}^{22}E_{j}^{21}],\no\\
&&M_{j-1,j}^{21}=\frac{i\eta}{\eta+\lambda}(E_{j-1}^{11}E_{j}^{12}+E_{j-1}^{12}E_{j}^{22}
+E_{j-1}^{33}E_{j}^{12})
+\frac{i\eta}{\eta-\lambda}[E_{j-1}^{12}(E_{j}^{11}+E_{j}^{33})+E_{j-1}^{22}E_{j}^{12}],\no\\
&&M_{j-1,j}^{23}=\frac{i\eta}{\eta+\lambda}(E_{j-1}^{33}E_{j}^{32}+E_{j-1}^{32}E_{j}^{22}
+E_{j-1}^{11}E_{j}^{32})
+\frac{i\eta}{\eta-\lambda}[E_{j-1}^{32}(E_{j}^{33}+E_{j}^{11})+E_{j-1}^{22}E_{j}^{32}],\no\\
&&M_{j-1,j}^{32}=\frac{i\eta}{\eta-\lambda}(E_{j-1}^{33}E_{j}^{23}+E_{j-1}^{23}E_{j}^{22}
+E_{j-1}^{11}E_{j}^{23})
+\frac{i\eta}{\eta+\lambda}[E_{j-1}^{23}(E_{j}^{33}+E_{j}^{11})+E_{j-1}^{22}E_{j}^{23}],\no\\
&&M_{j-1,j}^{13}=\frac{i\lambda}{\eta+\lambda}E_{j-1}^{21}E_{j}^{32}
-\frac{i\lambda}{\eta-\lambda}E_{j-1}^{32}E_{j}^{21},\no\\
&&M_{j-1,j}^{31}=\frac{i\lambda}{\eta+\lambda}E_{j-1}^{23}E_{j}^{12}
-\frac{i\lambda}{\eta-\lambda}E_{j-1}^{12}E_{j}^{23}.
\end{eqnarray}
From the equations (\ref{Lax-bulk}) and (\ref{Lax-boundary}), it follows that the $M_-(\lambda)$ matrix\footnote{The matrix $M_-(\lambda)$ acts non-trivially on spaces $0$, and $1$.} has the following form
\begin{eqnarray}
M_-(\lambda)=\left(
\begin{array}{ccc|ccc|ccc}
  D_1^{(1)}&   &   &   &   &   &   &   &   \\
          &D_1^{(2)}&   &A_-&   &   &   &   &   \\
          &   &D_1^{(3)}&   &   &   &   &   &   \\
          \hline
          &A_+&   &D_1^{(4)}&   &   &   &   &   \\
          &   &   &   &D_1^{(5)}&   &   &   &   \\
          &   &   &   &   &D_1^{(6)}&   &C_+&   \\
          \hline
          &   &   &   &   &   &D_1^{(7)}&   &   \\
          &   &   &   &   &C_-&   &D_1^{(8)}&   \\
          &   &   &   &   &   &   &   &D_1^{(9)}\\
      \end{array}
    \right),
\label{M-}\end{eqnarray}
with
\begin{eqnarray}
&&D_1^{(1)}=D_1^{(5)}=D_1^{(9)}=\frac{i \eta^2}{\eta^2 - \lambda^2},\,\no\\
&&D_1^{(2)}=\frac{i (b_1 - a_1) \eta^2}{\eta^2 - \lambda^2},\,D_1^{(4)}=\frac{i (a_1 - b_1) \eta^2}{\eta^2 - \lambda^2},\,
D_1^{(6)}=\frac{i (c_1 - b_1) \eta^2}{\eta^2 - \lambda^2},\,
D_1^{(8)}=\frac{i (b_1 - c_1) \eta^2}{\eta^2 - \lambda^2},\quad\no\\
&&D_1^{(3)}=\frac{i (c_1- a_1+1 ) \eta^2 + i(a_1 - c_1) \lambda^2}{\eta^2 - \lambda^2},\quad
D_1^{(7)}=\frac{i (a_1- c_1+1 ) \eta^2 + i(c_1 - a_1) \lambda^2}{\eta^2 - \lambda^2},\no\\
&&A_-=\frac{i \eta (\eta + (b_1-a_1) \lambda)}{\eta^2 - \lambda^2},\quad
A_+=\frac{i \eta (\eta + (a_1-b_1) \lambda)}{\eta^2 - \lambda^2},\no\\
&&C_-=\frac{i \eta (\eta + (b_1-c_1) \lambda)}{\eta^2 - \lambda^2},\quad
C_+=\frac{i \eta (\eta + (c_1-b_1) \lambda)}{\eta^2 - \lambda^2}.
\end{eqnarray}

The $M_+(\lambda)$ matrix\footnote{The matrix $M_+(\lambda)$ acts non-trivially on spaces $0$, and $N$.} has the following form

\begin{eqnarray}
M_+(\lambda)=\left(
\begin{array}{ccc|ccc|ccc}
  D_N^{(1)}&   &   &   &   &   &   &   &   \\
          &D_N^{(2)}&   &A'_+&   &   &   &   &   \\
          &   &D_N^{(3)}&   &   &   &   &   &   \\
          \hline
          &A'_-&   &D_N^{(4)}&   &   &   &   &   \\
          &   &   &   &D_N^{(5)}&   &   &   &   \\
          &   &   &   &   &D_N^{(6)}&   &C'_-&   \\
          \hline
          &   &   &   &   &   &D_N^{(7)}&   &   \\
          &   &   &   &   &C'_+&   &D_N^{(8)}&   \\
          &   &   &   &   &   &   &   &D_N^{(9)}\\
      \end{array}
    \right),
\label{M+}\end{eqnarray}
with
\begin{eqnarray}
&&D_N^{(1)}=D_N^{(5)}=D_N^{(9)}=\frac{i \eta^2}{\eta^2 - \lambda^2},\quad D_N^{(2)}=\frac{i (b_N - a_N) \eta^2}{\eta^2 - \lambda^2},\,\no\\
&&D_N^{(4)}=\frac{i (a_N - b_N) \eta^2}{\eta^2 - \lambda^2},\,
D_N^{(6)}=\frac{i (c_N - b_N) \eta^2}{\eta^2 - \lambda^2},\,
D_N^{(8)}=\frac{i (b_N - c_N) \eta^2}{\eta^2 - \lambda^2},\quad\no\\
&&D_N^{(3)}=\frac{i (c_N- a_N+1 ) \eta^2 + i(a_N - c_N) \lambda^2}{\eta^2 - \lambda^2},\no\\
&&D_N^{(7)}=\frac{i (a_N- c_N+1 ) \eta^2 + i(c_N - a_N) \lambda^2}{\eta^2 - \lambda^2},\no\\
&&A'_-=\frac{i \eta (\eta + (b_N-a_N) \lambda)}{\eta^2 - \lambda^2},\quad
A'_+=\frac{i \eta (\eta + (a_N-b_N) \lambda)}{\eta^2 - \lambda^2},\no\\
&&C'_-=\frac{i \eta (\eta + (b_N-c_N) \lambda)}{\eta^2 - \lambda^2},\quad
C'_+=\frac{i \eta (\eta + (c_N-b_N) \lambda)}{\eta^2 - \lambda^2}.
\end{eqnarray}
Thus we have obtained all the Lax pair operators for the free Shor-Movassagh spin chain with open boundaries. This also gives a straightforward proof of the integrability of the model.

\section{Boundary reflection matrices}
\setcounter{equation}{0}
We are now ready to compute the boundary $K$-matrices.
We proceed to study the constraint conditions (\ref{K-reflection}) and (\ref{K+reflection}). Let us set
\begin{eqnarray}
K_-(\lambda)=\left(
\begin{array}{ccc}
K_-^{11}&   &   \\
     &K_-^{22}&   \\
     &   &K_-^{33}\\
      \end{array}
    \right).
\end{eqnarray}
Substituting (\ref{M-}) into (\ref{K-reflection}), after tedious calculations, we have the following constraints for $K_-$,
\begin{eqnarray}
\frac{K_-^{11}}{K_-^{22}}=\frac{-(a_1-b_1)\lambda+\eta}{(a_1-b_1)\lambda+\eta},\qquad
\frac{K_-^{33}}{K_-^{22}}=\frac{-(c_1-b_1)\lambda+\eta}{(c_1-b_1)\lambda+\eta}.
\end{eqnarray}
Similarly, for $K_+(\lambda)$
\begin{eqnarray}
K_+(\lambda)=\left(
\begin{array}{ccc}
K_+^{11}&   &   \\
     &K_+^{22}&   \\
     &   &K_+^{33}\\
      \end{array}
    \right).
\end{eqnarray}
In order to calculate $K_+(\lambda)$, one should note that for
\begin{eqnarray}
\mathbb{U}_N(\lambda)=L_N(\lambda)\mathbb{U}_{N-1}(\lambda)L_N^{-1}(-\lambda),
\end{eqnarray}
the entries of $\mathbb{U}_{N-1}(\lambda)$ commute with the elements of $L_N(\lambda)$ (since they act on different quantum spaces). This fact simplifies the calculations of partial trace procedure in equation (\ref{K+reflection}). It also implies that the constraints of $K_+$ do not rely on $K_-$.

Substituting (\ref{M+}) into (\ref{K+reflection}), we have the constraints for entries of $K_+$,
\begin{eqnarray}
&&K_+^{11}[(a_N-b_N+1)\eta+(a_N-b_N)\lambda]+K_+^{22}[(a_N-b_N-1)\eta+(a_N-b_N)\lambda]\no\\
&&~~+K_+^{33}[(a_N-b_N+1)\eta+(a_N-b_N)\lambda]=0,\no\\
&&K_+^{11}[(c_N-b_N+1)\eta+(c_N-b_N)\lambda]+K_+^{22}[(c_N-b_N-1)\eta+(c_N-b_N)\lambda]\no\\
&&~~+K_+^{33}[(c_N-b_N+1)\eta+(c_N-b_N)\lambda]=0.
\end{eqnarray}
The above constraint equations give the ratios of the elements in $K$-matrices. In principle, one can write down many $K$-matrices which make the model integrable.
To retrieve the Hamiltonian (\ref{Hamiltonian}), an obvious solution is corresponding to Two-Boundary Temperley-Lieb algebra. Taking into account the scaling of the $K$-matrices, and setting $a_1=c_1$, $a_N=c_N$, we have
\begin{eqnarray}
K_-(\lambda)=\left(
\begin{array}{ccc}
-(a_1-b_1)\lambda+\eta&   &   \\
     &(a_1-b_1)\lambda+\eta&   \\
     &   &-(a_1-b_1)\lambda+\eta\\
      \end{array}
    \right),
\end{eqnarray}
and
\begin{eqnarray}
K_+(\lambda)\hspace{-1mm}=\hspace{-1mm}{2\over3}\hspace{-1mm}\left(
\begin{array}{ccc}
\hspace{-1mm}(b_N-a_N)(\lambda+\eta)\hspace{-1mm}+\hspace{-1mm}\eta&   &   \\
     &2(a_N-b_N)(\lambda+\eta)\hspace{-1mm}+\hspace{-1mm}2\eta&   \\
     &   &(b_N-a_N)(\lambda+\eta)\hspace{-1mm}+\hspace{-1mm}\eta\hspace{-1mm}\\
      \end{array}
    \right).
\end{eqnarray}
Here $a_1$, $b_1$, $a_N$, and $b_N$ are arbitrary free parameters describing the boundary effect.
Unlike the case of the one dimensional Hubbard open chain \cite{Zhou96,Guan97}, in our case the $K$-matrices can only be derived from the Lax formulation. The reflection equations (solutions) \cite{Sklyanin88} do not exist due to the vanishing of the crossing symmetry of the $R$-matrix.

Then the free Shor-Movassagh Hamiltonian with open boundaries (\ref{Hamiltonian}) can be derived from the double row transfer matrix (\ref{Double-trans}) in the following way:
\begin{eqnarray}
H_{FSM,open}&=&-{\eta\over2}\;\frac{\partial\ln \tau(\lambda)}{\partial \lambda}\bigg{|}_{\lambda=0}+\text{const.}
\end{eqnarray}
This is similar to the situation of the one dimensional Heisenberg spin chain.

\section{Conclusion}
We have presented the Lax pair for the free Shor-Movassagh open spin chain.
The double row monodromy matrix and transfer matrix of the spin chain have also been constructed.
We also found the boundary $K$-matrices. Note that unlike the case of Hubbard model \cite{Zhou96,Guan97,Guan2000} \footnote{The authors matched the results derived from both the Lax formulation and the reflection equations for Hubbard model.}, the $K$-matrices  can not be obtained from the reflection equations and Temperley-Lieb algebra.
We further remark that the $K_+$ and $K_-$ here are not isomorphic. The non diagonal open boundaries $K$-matrices solutions can be derived by assuming non diagonal boundary terms in the Hamiltonian (\ref{Hamiltonian}).
The integrability for open free Shor-Movassagh chain does not satisfy Sklyanin's reflection equations: the construction of  algebraic Bethe ansatz for the open chain will be hard.

\section*{Acknowledgments}

The work of K.H. was supported by the National Natural Science Foundation
of China (Grant Nos. 11805152, and 11947301) and Shaanxi Key Laboratory for Theoretical Physics Frontiers in China. K.H. would like to thank Prof. Wen-Li Yang for the helpful discussions.
V.K. was supported by SUNY center for QIS at Long Island project number CSP181035.
The authors would like to thank the Simons Center for Geometry and Physics for hospitality.
This work was started during the workshop
``Entanglement and Dynamical Systems: December 10-14, 2018" held at the center.

\end{document}